# A New Noise-Assistant LMS Algorithm for Preventing the Stalling Effect

**Hamid Reza Shahdoosti, Department of Electrical Engineering, Hamedan University of Technology**

**Email: h.doosti@hut.ac.ir**

**Abstract**

In this paper, we introduce a new algorithm to deal with the stalling effect in the LMS algorithm used in adaptive filters. We modify the update rule of the tap weight vectors by adding noise, generated by a noise generator. The properties of the proposed method are investigated by two novel theorems. As it is shown, the resulting algorithm, called Added Noise LMS (AN-LMS), improves the resistance capability of the conventional LMS algorithm against the stalling effect. The probability of update with additive white Gaussian noise is calculated in the paper. Convergence of the proposed method is investigated and it is proved that the rate of convergence of the introduced method is equal to that of LMS algorithm in the expected value sense, provided that the distribution of the added noise is uniform. Finally, it is shown that the order of complexity of the proposed algorithm is linear as the conventional LMS algorithm.

*Keywords*: Adaptive filter, LMS algorithm, stalling effect, finite precision effect.

# 1. Background

Adaptive filter algorithms are widely used for channel estimation and equalization in digital communication systems and digital signal processing [1-2]. Particularly, Image fusion [3-6] and image denoising [7-10] are two different tasks which can be performed by adaptive filters. The performance analysis of adaptive filter algorithms is usually done based on analog assumptions in infinite precision environments. In practice, digital signal processors are used to implement these algorithms. Using these processors updating the filter tap weights, calculation of the estimation error and data sampling are done in a finite precision environment. This finite precision assumption brings about some phenomena. One of these phenomena is the quantization error which takes place in converting analog data to digital ones. In can be shown that it is possible to consider the quantization noise as an additive independent source of white noise provided that the quantization is performed with high resolution (using 6 bits or larger) and also the signal spectra is sufficiently rich [11-12]. One of the challenges in the implementation of adaptive filter algorithms in finite precision environments is the stalling effect. In a finite precision environment e.g., a processor, whenever the correction term for a specific tap weight is smaller in magnitude than the half of the least significant bit ( LSB) of this tap weight, the corresponding tap weight in the algorithm is not updated (according to the rule of rounding to the nearest mode) and thus, this filter tap weight stalls [13]. In order to prevent the stalling effect in a finite precision environment from happening, the residual error should be made as small as possible. For this purpose, one of the following two methods is usually used [12]:

1) Using a large number of bits for representing the filter tap weight and other data by which the LSB can be reduced.
2) The step-size parameter may be made as large as possible in such a way that the convergence of the algorithm is still guaranteed.

Another method for combating the stalling effect, is using *dither* in the quantizer input by which the tap weight accumulator is fed [14]. The authors in [15] modeled the coefficients of adaptive filter as a Markov chain and the matrix of transition probabilities of the chain was determined for the one-dimensional case in this model. In addition, the conditions in which stalling phenomenon occurs, was determined in [15]. In [16] a modification of the LMS algorithm was proposed that alleviates the effect of quantization at virtually no extra computational cost. In this algorithm, stalling situations are detected and a secondary adaptive filter is used to increase the precision in such situations. A method showing a performance that is comparable to that of full precision adaptive filters has been proposed in [17], which uses a companded delta modulation structure. In [18] the quantization effects on the steady-state performance of a fixed-point implementation of the LMS adaptive algorithm was studied, and the stall mode was reviewed. Furthermore, the value of step-size corresponding to the onset of the stall mode has been predicted in [18], such that one can avoid the stalling phenomenon by judiciously choosing the step size value.

In this paper we propose a new method which is capable of preventing the stalling effect by using a limited number of bits. The main contribution of this paper is that the algorithm does not stop updating even when the correction term is smaller in magnitude than the half of the LSB. The proposed algorithm has a rate of convergence almost equal to that of LMS algorithm. The rest of the paper is organized as follows. In section 2 the proposed algorithm is presented in details. Also, we analyze the rate of convergence of the proposed algorithm in this section. Section 3 presents the simulation results and comparisons for the proposed method and conventional LMS algorithm in finite and infinite precision environments. Section 4 concludes the essay.

## 2. Algorithm Statement

### 2.1. Preliminaries

The conventional LMS algorithm updates the tap weights as:

$$\boldsymbol{W}(i+1) = \boldsymbol{W}(i) + \mu \boldsymbol{u}(i) e(i) \tag{1}$$

where $\boldsymbol{u}(i) = [u(i),\dots,u(i-M+1)]^T$ and $\boldsymbol{W}(i) = [w_0(i),\dots,w_{M-1}(i)]^T$ are the input data vector and tap weight vector, respectively. Also $e(i) = d(i) - \hat{d}(i) = \boldsymbol{W}(i)\boldsymbol{u}(i) - d(i)$ is the estimation error.

Fig. 1 shows the block diagram for a general adaptive filtering algorithm. In the above algorithm, if the values of each element of the vector $\mu \boldsymbol{u}(i)e(i)$ in the finite precision environment is less than **0.5** $LSB$, the stalling effect happens and the value of that element is considered as zero. Therefore, that tap weight is not updated. Assume that our algorithm rounds any number to the nearest LSB. We consider the tap weight corresponding to the input $u(i)$. Define threshold error $e_T(i)$ as the minimum acceptable value of estimation error above which no stalling effect happens for the tap weight corresponding to input $u(i)$. We have $e_T(i) = \frac{LSB}{2\mu u(i)}$.

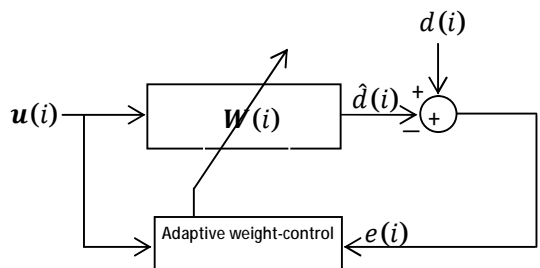

**Fig**. **1**. Block diagram of an adaptive transversal filter

If $e(i) < e_T(i)$ then $\mu u(i)e(i)$ is considered as zero. For example suppose that 12 bits are allocated for data representation in the decimal part. In this case $LSB = 2.44 \times 10^{-4}$. Assuming the normalized input data equals to **0.5** and $\mu = \mathbf{0.01}$, the threshold error is equal to $e_T(i) = \mathbf{0.0244}$. So the value of estimation error cannot exceed this threshold value.

Fig. 2 demonstrates how the stalling effect happens in this finite precision environment. The system model is an AR model described as $u(i) = 0.1u(i-1) + 0.1u(i-2) ... + 0.1u(i-10)$. The initial input data is considered as $u(1:10) = [0.2\ 0.3\ 0.28\ 0.26\ 0.4\ 0.24\ 0.46\ 0.6\ 0.56\ 0.48]$. Also we set the initial tap weight elements as **0.01**. As it can be seen from the figure, the finite precision LMS stops updating its tap weight vector when the value of error is less than **0.0244**. This begins at iteration number **134**.

Is it possible to reach less error values with the same number of bits allocated for digital quantization? We answer this question in the following section.

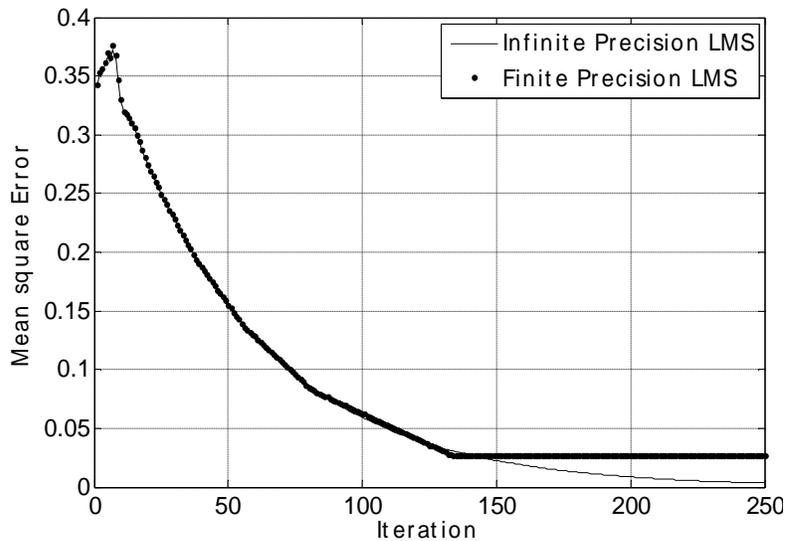

**Fig. 2**. Stalling effect in finite precision LMS.

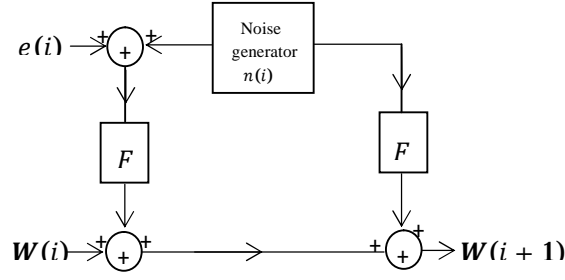

**Fig. 3**. Details of the adaptive weight-control mechanism in the AN-LMS algorithm ($F$ is a function defined as $F(x) = \mu u(i)x$).

## 2.2. Proposed Algorithm

In this section we present the proposed algorithm dealing with the stalling effect. Fig. 3 depicts the structure of Adaptive weight-control mechanism in the AN-LMS algorithm. Specifically how new weights are calculated from previous ones. Function $F$ is defined as $F(x) = \mu u(i)x$. The update equation for filter weights can be written as:

$$\boldsymbol{W}(i + 1) = \boldsymbol{W}(i) + \mu \boldsymbol{u}(i)\big(e(i) + n(i)\big) - \mu \boldsymbol{u}(i)n(i) \tag{2}$$

where $n(i)$ is a white noise in this formula. The algorithm is composed of two update terms i.e., $\mu \boldsymbol{u}(i)\big(e(i) + n(i)\big)$ and $\mu \boldsymbol{u}(i)n(i)$. Note that if the environment is not a finite precision one, then this algorithm is exactly the same as the LMS algorithm. This simple algorithm has some interesting properties. As we prove later, this algorithm prevents the stalling effect from happening. If the error value is bigger than the threshold value, i.e., $e(i) \geq e_T(i)$, then rates of convergence of this algorithm and the LMS algorithm are almost equal. This is true because the quantized version of modification terms $\mu \boldsymbol{u}(i)e(i)$ in the LMS algorithm and $\mu \boldsymbol{u}(i)\big(e(i) + n(i)\big) - \mu \boldsymbol{u}(i)n(i)$ in the proposed algorithm differ at most in $1\ LSB$ which $1\ LSB$ and $-1\ LSB$ take place with equal probability. If $e(i) < e_T(i)$ AN-LMS algorithm prevents the stalling effect. Assume that $e(i) < e_T(i) = \frac{LSB}{2\mu u(i)}$ in which the stalling occurs for the LMS algorithm. Suppose that the noise $n(i)$ has an arbitrary distribution (e.g. Gaussian). Also assume that the error value $e(i)$ and the input signal $u(i)$ are both positive. Since the value of error is less than $\frac{LSB}{2\mu u(i)}$ and error value is positive, $\frac{LSB}{2\mu u(i)} - e(i)$ is positive. If the added noise value falls within the interval which is shown in Fig. 4, then $\mu u(i)\big(e(i) + n(i)\big)$ in AN-LMS algorithm exceeds $\mu u(i)n(i)$ by $1\ LSB$ therefore the weight will be updated by $+1 LSB$ in the true direction i.e., in the direction of sign LMS algorithm. This interval is not the only interval within which if noise value falls, the weight will be updated.

In fact if added noise falls within the intervals $[\frac{(2k-1)LSB}{2\mu u(i)} - e(i), \frac{(2k-1)LSB}{2\mu u(i)})$ for some integer $k$, the weights will be updated in the true direction.

The following two theorems defined and proved by the authors, investigate the properties of the AN-LMS algorithm.

**Theorem 1**: Consider the algorithm presented by equation (2). Suppose that in the time instant $i$ we have $e(i) < \frac{LSB}{2u(i)\mu}$, i.e., the stalling phenomenon has happened in the finite precision environment for the LMS algorithm, then the tap weight $w_0(i)$ in the AN-LMS algorithm is updated in the true direction i.e., in the direction of the sign LMS algorithm.

*Proof-* Assume the added noise $n(i)$ has an arbitrary distribution (e.g. Gaussian). Also assume that the error value and input signal $u(i)$ are positive. Since the value of error is less than $\frac{LSB}{2\mu u(i)}$ and it is also positive, one can conclude that $\frac{LSB}{2\mu u(i)} - e(i)$ is positive. If the noise value falls within the interval $\frac{(2k-1)LSB}{2\mu u(i)} - e(i) \leq n(i) < \frac{(2k-1)LSB}{2\mu u(i)}$ for some integer $k$ then $\mu u(i)(e(i) + n(i)) - \mu u(i)n(i)$ will be equal to $1\ LSB$. So, the tap weight $w_0(i)$ will be updated by $1\ LSB$ in the true direction (like sign LMS algorithm). If the noise value does not fall in these intervals, then no updating in the tap weight will happen. We prove this in the following. Suppose that:

$$\frac{(2k-1)LSB}{2\mu u(i)} - e(i) \leq n(i) < \frac{(2k-1)LSB}{2\mu u(i)} \quad (3)$$

then,

$$\frac{(2k-1)LSB}{2\mu u(i)} \leq n(i) + e(i) < \frac{(2k-1)LSB}{2\mu u(i)} + e(i) \quad (4)$$

and thus,

$$\frac{(2k-1)LSB}{2} \leq \mu u(i)(n(i) + e(i)) < \frac{(2k-1)LSB}{2} + \mu u(i)e(i) \quad (5)$$

Since $\mu u(i)e(i) < 0.5 LSB$ the value of $\mu u(i)(n(i) + e(i))$ will be equal to $k\ LSB$.

Now we analyze the second update term of equation (2), i.e., $\mu u(i)n(i)$. We have

$$\frac{(2k-1)LSB}{2\mu u(i)} - e(i) \leq n(i) < \frac{(2k-1)LSB}{2\mu u(i)} \quad (6)$$

which is equivalent to

$$\frac{(2k-1)LSB}{2} - \mu u(i)e(i) \leq \mu u(i)n(i) < \frac{(2k-1)LSB}{2} \quad (7)$$

As it can be inferred from this equation, since $\mu u(i)e(i) < 0.5 LSB$ the value of $\mu u(i)n(i)$, when rounded to the nearest LSB, will be equal to $(k-1)LSB$ and thus the difference between two update terms is equal to $1LSB$ and the weight will be updated by $+1LSB$ in the true direction ($u(i)$ and $e(i)$ are both positive).

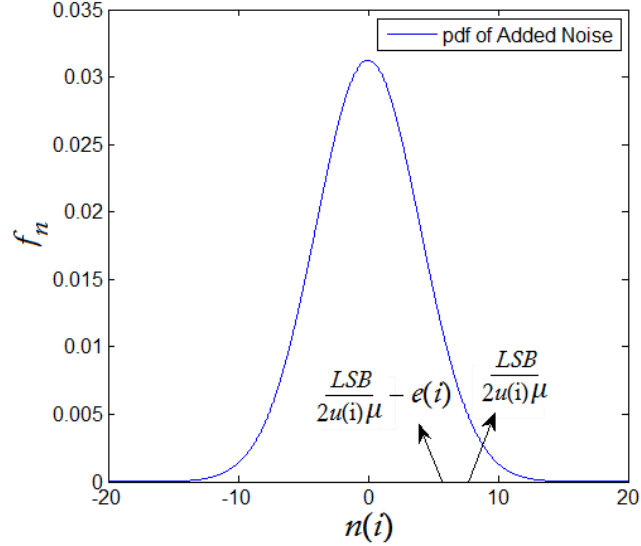

**Fig. 4.** Interval in which AN-LMS updates.

Now consider the case that the noise value does not fall within the previous interval but falls in the interval:

$$\frac{(2k-1)LSB}{2\mu u(i)} \leq n(i) < \frac{(2k+1)LSB}{2\mu u(i)} - e(i) \tag{8}$$

We show that in this case, the proposed algorithm does not update the tap weight. From (3) we have:

$$\frac{(2k-1)LSB}{2} + \mu u(i)e(i) \leq \mu u(i)(n(i) + e(i)) < \frac{(2k+1)LSB}{2} \tag{9}$$

Since $\mu u(i)e(i) < 0.5LSB$ the value of $\mu u(i)(n(i) + e(i))$, when rounded to the nearest LSB, will be equal to $kLSB$. Moreover, from (3) we can write:

$$\frac{(2k-1)LSB}{2} \leq \mu u(i)n(i) < \frac{(2k+1)LSB}{2} - \mu u(i)e(i) \tag{10}$$

In similar way, the value of $\mu u(i)n(i)$, when rounded to the nearest LSB, will be equal to $k\,LSB$. Therefore, the difference between two update terms is equal to zero and the tap weight is not updated by the proposed algorithm. This result can be concluded by the similar expressions for other tap weights $w_1(i)$ to $w_{M-1}(i)$. In addition, the proof for other 3 possible states for signs of $u(i)$ and $e(i)$ is similar (the intervals changes according to signs of $u(i)$ and $e(i)$). The proof is completed here.

**Theorem 2**: Consider the algorithm proposed in (1) and also suppose that the distribution of the added noise $n(i)$ is uniform. Rate of convergence of the proposed algorithm and that of the LMS algorithm is identical in the expected value sense.

*Proof-* In the LMS algorithm the update term is $\mu u(i)e(i)$. If $\mu u(i)e(i) \geq 0.5LSB$ the probability that the update term in the AN-LMS algorithm exceeds that of the LMS algorithm by $1LSB$ is equal to the probability that the update term in the LMS algorithm exceeds that of the AN-LMS algorithm by $1LSB$. Therefore, the rates of convergence of AN-LMS and LMS are equal in a finite precision environment, provided that the stalling phenomenon has not occurred and can be approximated by a single exponential curve. An average eigenvalue can be defined for the underlying correlation matrix $R$ of the tap inputs as:

$$\lambda_{av} = \frac{1}{M}\sum_{j=1}^{M}\lambda_j \tag{11}$$

The learning curve of the LMS algorithm can be approximated by a single exponential with time constant $\tau$. One can use equations which are developed for the method of steepest descent, to define average time constant for the LMS algorithm as the following [12]:

$$\tau = \frac{1}{2\mu\lambda_{av}} \tag{12}$$

On the other hand, if $\mu u(i)e(i) < 0.5LSB$, as it is stated before, the algorithm always updates the tap weights in the true direction. Now, assume that the distribution of the added noise is uniform. We prove that in the case $\mu u(i)e(i) < 0.5LS$, the rates of convergence of the proposed method and the LMS algorithm are equal. Again assume that the error value $e(i)$ and input signal $u(i)$ may be negative or positive.

Define $A$ as the event that the noise value falls within the intervals:

$$[\frac{(2k-1)LSB}{2\mu u(i)} - e(i), \frac{(2k-1)LSB}{2\mu u(i)}) \quad k \in (-\infty, +\infty) \tag{13}$$

Suppose that $\Pr(A) = p$. As it is proved in theorem 1, the AN-LMS algorithm updates the tap weights if and only if $A$ happens. Then the expected value of the update term in the proposed method is equal to:

$$E(w_0(i+1) - w_0(i)) = p \times 1LSB \times sign(u(i)e(i)) + (1-p) \times 0 \tag{14}$$

which $w_0(i)$ is tap weight corresponding to the input $u(i)$.

Now we calculate $p$ as:

$$p = \left|\frac{e(i)}{\frac{LSB}{\mu u(i)}}\right| \tag{15}$$

So the expected value of the update term in the proposed method is:

$$E(w_0(i+1) - w_0(i)) = \left|\frac{e(i)}{\frac{LSB}{\mu u(i)}}\right| \times 1LSB \times sign(u(i)e(i)) = \mu u(i)e(i) \tag{16}$$

which means that the algorithm has the same rate of convergence as the LMS algorithm in the expected value sense. So the approximation of learning curve with an exponential function is still valid when the stalling phenomenon occurs. This completes the proof. This equality in the expected value sense can be observed also in the simulation results which is presented in section 3.

### 2.3. Convergence of the AN-LMS algorithm

As mentioned, this algorithm is the same as LMS algorithm in infinite precision environments. We have modified the LMS algorithm by adding and subtracting a noise produced by the noise generator (see equation (2)). So, the convergence of AN-LMS algorithm is the same as that of LMS algorithm in an infinite precision environment. The transient component of the mean squared-error $J(i)$ dies out, which means that the LMS algorithm and AN-LMS algorithm are convergent in the mean square if and only if the step-size parameter $\mu$ satisfies the following condition [19]:

$$0 < \mu < \frac{2}{\lambda_{Max}} \qquad (17)$$

where $\lambda_{Max}$ is the largest eigenvalue of the correlation matrix of input data, i.e., $R$.

The simplest consistent estimator for $R$ can be obtained by using instantaneous estimates that are based on sample values of the tap input vector as the following:

$$\tilde{R} = \boldsymbol{u}(i) * \boldsymbol{u}(i)^H \qquad (18)$$

The condition for the LMS algorithm and AN-LMS algorithm to be convergent in the mean square, which is described in Eq. (17), needs a knowledge of the largest eigenvalue, $\lambda_{Max}$, of the described correlation matrix R. In the application of these algorithms, knowledge of $\lambda_{Max}$ is not usually available. To deal with this practical difficulty, the trace of R may be considered as a conservative estimate for $\lambda_{Max}$. Therefore, the condition described in Eq. (17) can be reformulated as:

$$0 < \mu < \frac{2}{tr[R]} \qquad (19)$$

### 2.4. Complexity of the AN-LMS Algorithm

We consider the complexity as the number of multiplications needed to calculate the updated tap weight vector from the previous one. The computational complexity for the conventional LMS is equal to $2M$ and thus is linear with the number of taps i.e. $O(M)$. Since $\mu u(i)$ exists in both update terms of the AN-LMS algorithm, the AN-LMS algorithm has one additional multiplication. Therefore its complexity is equal to $3M$. Thus, the AN-LMS algorithm improves the resistance to the stalling effect while keeping the complexity linear with the number of taps i.e. $O(M)$.

## 2.5. Probability of Update with Additive Gaussian Noise

In order to calculate the probability of update in the AN-LMS algorithm knowledge of the distribution of added noise is necessary. Assume the added noise has a zero mean Gaussian distribution with variance $\sigma^2$. We proved that if the noise value falls within the interval $\frac{(2k-1)LSB}{2\mu u(i)} - e(i) \leq n(i) < \frac{(2k-1)LSB}{2\mu u(i)}$ for some integer $k$ then AN-LMS updates the tap weights. Suppose that $l = \left|\frac{LSB}{2\mu u(i)}\right|$ and $e(i)$ is positive. The probability of update event can be written as:

$$p = \sum_{k=-\infty}^{\infty}[\varphi\left(\frac{(2k-1)l}{\sigma}\right) - \varphi\left(\frac{(2k-1)-e}{\sigma}\right)] \quad (20)$$

where $\varphi(x)$ is the cumulative distribution function (CDF) of the standard normal distribution. Now we evaluate the value of **P**. In the first scenario, we suppose that $e(i)$ is constant and $l$ and $\sigma$ are variables. The result of simulation has been shown in Fig. 5.

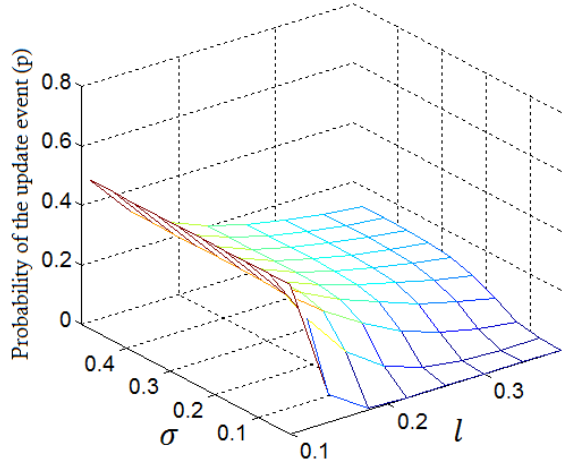

**Fig. 5**. Probability of the update event as function of $l$ and $\sigma$ when $e(i)$ is constant.

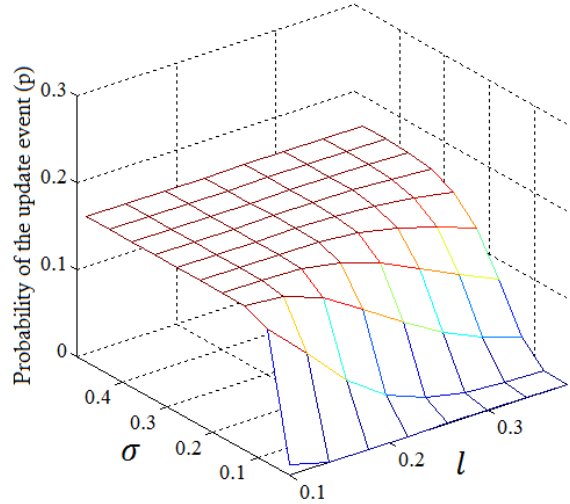

**Fig. 6**. Probability of update as function of $l$ and $\sigma$ when $e(i) = \frac{l}{3}$.

As it is depicted in Fig. 5, the value of $p$ decreases with the increase of $l$ and increases with the increase of $\sigma$. In the second scenario we assume that $e(i) = \frac{l}{3}$ and $l$ and $\sigma$ are variable. The result has been shown in Fig. 6.

We can see from Fig. 6 that the value of $p$ decreases with the increase of $l$ and increases with the increase of $\sigma$. Also it can be seen that with increase of $\sigma$ the value of $p$ closes to a constant value $\frac{e}{2*l}$ which is equal to the probability of update event when the added noise has the uniform distribution. This result is expected because when $\sigma$ increases, the Gaussian distribution closes to the uniform distribution. So we can conclude that the uniform distribution has a faster rate of convergence than the Gaussian distribution.

## 3. Simulation Results

In this section we present the simulation results for AN-LMS algorithm.

Fig. 7 shows the mean absolute error for the system corresponding to Fig. 4. The two used adaptive filter algorithms are LMS and AN-LMS. The noise generator block in Fig. 2 generates the Gaussian noise. Fig. 7 shows the main feature of AN-LMS. In fact, the infinite precision LMS has no stalling effect problem. As it can be seen the AN-LMS does continue updating its tap weight vector. Also the rate of convergence of our method is almost equal to that of infinite precision LMS as it can be seen from the coincidence of the two curves. Note that the distribution of the noise is Gaussian but not uniform, thus the rates of convergence are not completely equal in expected value sense. In Figs. 4 and 7 the channel adds no noise to the input signal i.e. the channel impulse response is $h(n) = \delta(n)$.

Now we present the simulation results for a noisy channel which is a real case. Consider an AR signal described as

$$u(i) = 0.1u(i-1) + 0.1u(i-2) \ldots + 0.1u(i-11) + w(i) \quad (21)$$

where $w(i)$ is a white noise with variance **0.004**. Again the number of bits used for representation of the decimal part is **12**. Fig. 8 demonstrates that the AN-LMS attains lower estimation errors than the conventional finite precision LMS algorithm. Unlike the finite precision LMS the AN-LMS does not stall.

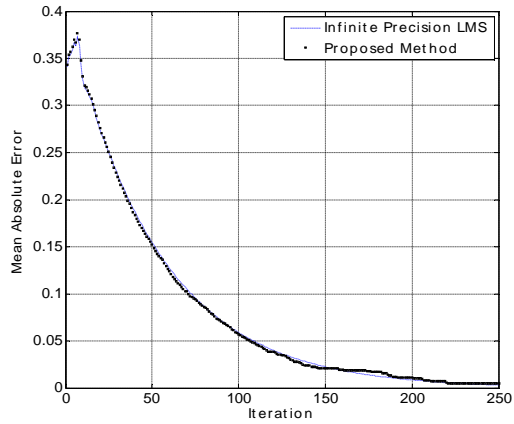

**Fig. 7**. Stalling effect in finite precision LMS.

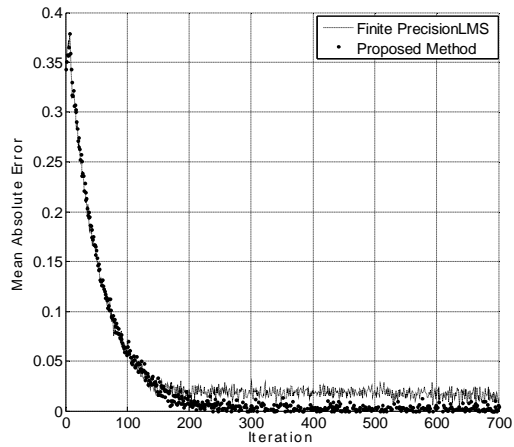

**Fig. 8**. Comparison between Finite Precision LMS and Proposed method in for an AR signal; the proposed method combats the stalling effect.

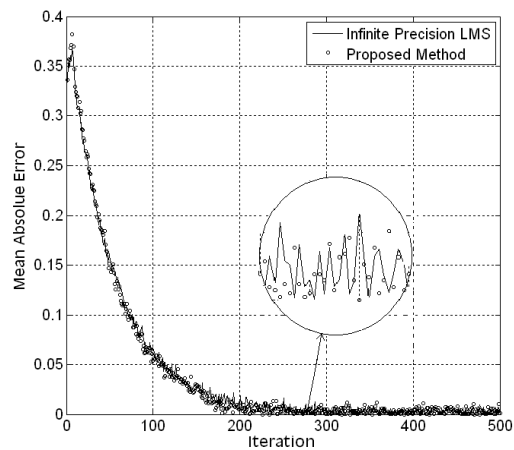

**Fig. 9**. Equality of rates of convergence for two adaptive filter algorithms.

Finally Fig. 9 compares the rates of convergences of the proposed method (AN-LMS) and the infinite LMS algorithm. Again the system used in Fig. 8, is utilized. The coincidence of the two learning curves suggests that the rates of convergence of the AN-LMS and the infinite precision LMS are equal in expected value sense.

## 4. Concluding Remarks

In this paper we presented an updated version of LMS algorithm, called AN-LMS algorithm. The basic difference of AN-LMS and LMS is that AN-LMS injects a noise to the structure of LMS. This makes AN-LMS algorithm have two update terms. We proved that in the situation that the stalling effect happens in conventional LMS, AN-LMS still updates its tap weights. The simulation results confirm this analytical proof. Another observation is that if the distribution of the added noise tends to uniform then the rate of converge of AN-LMS tends to that of infinite LMS. Again, this is observed in simulation results too. Note that despite increasing the computational complexity, the AN-LMS algorithm has the same linear complexity order as the conventional LMS algorithm.